\documentclass{elsart}
\usepackage{amssymb}
\usepackage{amsmath}
\usepackage[dvips]{graphicx}
\journal{\textbf{International Journal of Non-Linear Mechanics\ vol. 47, p.p. 77-84 (2012).\qquad\qquad }}
\begin{document}
\def\func#1{\mathop{\rm #1}}%

\begin{frontmatter}

\title{ { Travelling waves near a critical point\\ of a binary fluid mixture}}

\author{Henri Gouin}
\ead{henri.gouin@univ-cezanne.fr}
 \address { C.N.R.S. U.M.R. 6181 \&
 University of Aix-Marseille, \\ Case 322, Av. Escadrille
 Normandie-Niemen, 13397 Marseille Cedex 20 France}
\author{Augusto Muracchini and Tommaso Ruggeri}
\ead{augusto.muracchini@unibo.it}
\address {Department of Mathematics and Research Center of Applied Mathematics \\
University of Bologna, Via Saragozza 8, 40123 Bologna, Italy.}
\ead{tommaso.ruggeri@unibo.it}

\begin{abstract}
Travelling waves of densities of   binary fluid mixtures are
investigated near a critical point. The free energy is considered in
a non-local form taking account of the density gradients.  The equations of motions are
applied to a universal form of the free energy near critical conditions and can be integrated
 by a rescaling process where the binary mixture is similar to a single fluid.
Nevertheless, density solution profiles obtained are not necessarily monotonic. As indicated in Appendix, the results might be extended to other topics like  finance or biology.
\end{abstract}

\end{frontmatter}

\section{Introduction}

In physical chemistry, thermodynamics and condensed matter physics, a
critical point specifies the conditions (temperature, pressure and
concentration) at  which distinct  phases do not exist  \cite{Bruhat,Groot,Levitch}%
. There are multiple types of critical points such as vapor-liquid or
liquid-liquid critical point. A single fluid has a unique critical point
associated with given temperature, pressure and density. For binary mixtures
of fluids, in the space of temperature, pressure, concentration, critical
points are represented by a curve in a convenient domain \cite{Scott}; to
each temperature we can associate a critical pressure and two critical
densities corresponding to the mixture components \cite{Emschwiller,Rocard}.
\newline
An important thermodynamical potential is related to the mixture volume free
energy \cite{Muller1,Muller2,ET}. At a given temperature, the volume free
energy is associated with the \emph{spinodal} curve connecting the two
different phases of the binary mixture. Due to conditions of equilibrium of
phases, it is possible to form a general expansion of the free energy near a
critical point. This form is known in the literature by means of physical
chemistry considerations \cite{Rowlinson1} and is the form we use in our calculations.

By calculations in molecular theories, the densities of the components
fluctuate near a critical point \cite{Israel}. In the following, we use a
continuous model to investigate how the average variations of densities are
related to molecular interactions. Two assumptions are explicit \cite%
{Bongiorno,Gouin7,Rowlinson2}:\newline
$i)$ \ The component densities are assumed to be smooth functions of the
distance from an interface layer which is assumed to be flat on the scale of
molecular sizes. The correlation lengths are assumed to be greater than
intermolecular distances  \cite{Hagan,Hohenberg}; this is the case when at a given temperature $T$
the parameters are close to the ones of a critical state \cite{Ono}.\newline
$ii)$ \ The binary mixture is considered in the framework of a mean-field
theory. This means, in particular, that the free energy of the mixture is a
classical so-called "gradient square functional". This kind of
Landau-Ginzburg model consisting of a quadratic form of the density
gradients comes from Maxwell and van der Walls original ideas \cite%
{Korteweg,Maxwell,vdW,Widom}. At given critical conditions, the coefficients
of the quadratic form are constant.

This point of view that, in non-homogeneous regions, the mixture may be
treated as bulk phase with a local free-energy density and an additional
contribution arising from the non-uniformity which may be approximated by a
gradient expansion truncated at the second order is most likely to be
successful and perhaps even quantitatively accurate near a critical point
\cite{Rowlinson1}. The approximation of mean field theory does provide a
good understanding and allows one to  explicitly calculate the magnitude of the
coefficients of the model. These non-linear equations are able to
represent interface layer and bulks and consequently allow to build a
complete theory of the mixtures in non-homogeneous domains in dynamics.

In Section 2 we recall the equations of motion in a pure mechanical
process obtained through the Hamilton variational principle.

Section 3 is devoted to travelling waves without dissipation. Due to the fact that the equations
are Galilean invariant, the case of equilibrium and the case of motion are analyzed  together.

In Section 4 by means of a rescaling process taking the vicinity of a
critical point into account, we integrate the equation for equilibrium as
well as for motions with dissipation.

Two appendices present the motion equations and the mathematical reason of
the choice of the free energy form near a critical point of a binary mixture
of fluids  obtained by a new method issued from differential geometry.

\section{Isothermal motion of a binary fluid mixture near a critical point}

We study a mixture of two fluids by a mechanical process. No assumption has
to be done about composition or miscibility. The motion of a two-fluid
continuum can be represented with two surjective differentiable mappings
(see Fig. 1) \cite{Bedford,Germain,Gouin4,Herivel}:
\begin{equation*}
\mathbf{z}\rightarrow \mathbf{X}_{1}=\Phi _{1}(\mathbf{z})\ \ \hbox{\rm and}%
\ \ \mathbf{z}\rightarrow \mathbf{X}_{2}=\Phi _{2}(\mathbf{z}) ,
\end{equation*}%
where subscripts 1 and 2 are associated with each constituent of the
mixture. Term $\mathbf{z}= (t,\mathbf{x})$ denotes Euler variables in
space-time $\mathcal{W}$ and terms $\mathbf{X}_{1}$ and $\mathbf{X}_{2}$ denote the
Lagrange variables of constituents in \emph{reference spaces} $\mathcal{D}%
_{o1}$ and $\mathcal{D}_{o2}$ respectively.
\begin{figure}[h]
\begin{center}
\includegraphics[width=7cm]{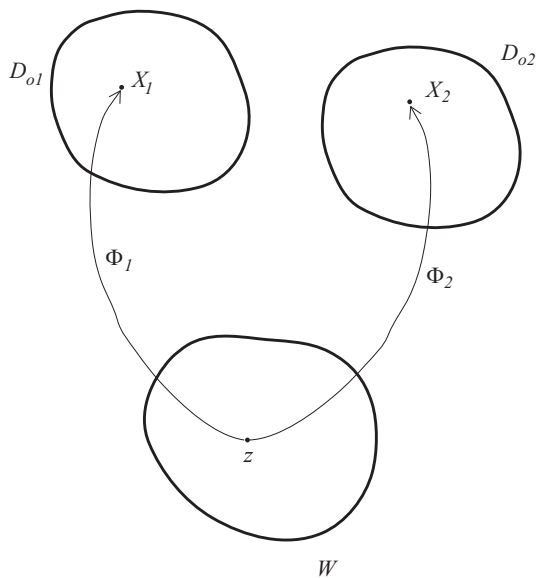}
\end{center}
\caption{{General representation of a two-fluid continuum motion.}}
\label{Fig. 1}
\end{figure}

In the pure mechanical case, the Lagrangian density of the mixture is
\begin{equation*}
L={\frac{1}{2}}\ \rho _{1}\mathbf{v}_{1}^{2}+{\frac{1}{2}}\ \rho _{2}\mathbf{%
v}_{2}^{2}-e -\rho _{1}\Omega _{1}-\rho _{2}\Omega _{2},
\end{equation*}%
where $\mathbf{v}_{1}$ and $\mathbf{v}_{2}$ denote the velocity vectors of
each constituent, $\rho _{1}$ and $\rho _{2}$ are the densities, $\Omega
_{1} $ and $\Omega _{2}$ are the external force potentials depending only on
$\mathbf{z}=(t,\mathbf{x})$ and $e$ is the volume energy \cite%
{Berdichevsky,Galdi}.\newline
The expression of the Lagrangian is in a general form. In fact dissipative
phenomena imply that $\mathbf{v}_{1}$ is almost equal to $\mathbf{v}_{2}$;
it is the reason why we do not take account of some kinetic energy
associated with the relative velocity of the components which is of
smaller order (at least of order 2) and will be negligible in travelling
wave behavior \cite{Bowen,Joseph,Landau}. Because of the interaction between
the constituents, the volume energy $e$ is not the sum of the energies of  each
constituent of the mixture, like for \textsl{Euler mixtures of fluids}. The
mixture is assumed not to be chemically reacting. Conservations of masses
require
\begin{equation}
\rho _{i}\,\hbox {det}\ \mathbf{F}_{i}=\rho _{oi}\ (\mathbf{X}_{i}),
\label{mass1}
\end{equation}%
where subscript $i$ belongs to $\{$1,2$\}$. At $t$ fixed, the deformation
gradient $\partial \mathbf{x}_i/\partial\mathbf{X}_i$ associated with $\Phi _{i}$ is denoted by $\mathbf{F}_{i}$ and $%
\rho _{oi}$ is the reference specific mass in $\mathcal{D}_{oi}$. \newline
Eqs (\ref{mass1}) are equivalent to the Eulerian form
\begin{equation}
{\frac{\partial \rho _{i}}{\partial t}}+\hbox{div}\ \rho _{i}\,\mathbf{v}%
_{i}=0.  \label{mass2}
\end{equation}%
The volume energy $e$ is given by the behavior of the mixture \cite%
{Cahn,Dunn2,Eglit}. In our mechanical case, for an energy depending on
gradients of densities, the volume energy is
\begin{equation*}
e =e (\rho _{1},\rho _{2},\hbox{grad}\rho _{1},\hbox{grad}\rho _{2}) .
\end{equation*}%
The potential
\begin{equation*}
\mu _{i}=\frac{\partial e }{\partial \rho _{i}}-\frac{\partial }{\partial
x_{\gamma }}\left( \frac{\partial e }{\partial \rho _{i,\gamma }}\right)
\end{equation*}%
defines the \emph{specific free enthalpy or chemical potential }of the
constituent $i$ of the mixture \cite{Gouin4}. Subscript $\gamma $
corresponds to the spatial derivatives associated with gradient terms.
Usually, summation is made on repeated subscript $\gamma $. In practice, we
consider a quadratic form with constant coefficients $C_{1},C_{2},D$\,
\begin{equation*}
Q=C_{1}\,(\hbox{grad}\,\rho_{1})^{2}+2D\,\hbox{grad}\rho _{1}\,\hbox{grad}%
\rho_{2}+C_{2}\,(\hbox{grad}\,\rho _{2})^{2}
\end{equation*}
such that
\begin{equation}
e =g_{o}(\rho _{1},\rho _{2})+{\frac{1}{2}}\ Q,  \label{nonlocal energy}
\end{equation}%
where $g_{o}(\rho _{1},\rho _{2})$ is the value of the volume energy of the
homogeneous bulks.

To obtain the equations of motions, we use a variational principle whose
original feature is to choice variations in reference spaces (Fig. 1). They
are associated with a two-parameter family of virtual motions of the mixture
(see Appendix A). \newline
The equation of the motion of each constituent of the mixture writes (\cite%
{Gouin-Ruggeri} and the references therein)
\begin{equation}
\mathbf{a}_{i}+\hbox {grad}\,(\mu _{i}+\Omega _{i})=0,\quad i=\{1,2\},
\label{motions 1}
\end{equation}%
where $\mathbf{a}_{i}$ denotes the acceleration of the component $i$
($i$=1,2). In applications, the motions are supposed to be isothermal ($T$
denotes the common temperature value of the two components) and correspond
to strong heat exchange between components. In thermodynamics, this case
corresponds to a function $g_{o}(\rho _{1},\rho _{2})$ as the volume free
energy of the homogeneous mixture at temperature $T$. \newline
In our model, the equations of motion (\ref{motions 1}) yield
\begin{equation}
\left\{
\begin{array}{l}
\mathbf{a}_{1}=\hbox {grad}\,\{C_{1}\,\Delta \rho _{1}+D\,\Delta \rho
_{2}-g\,_{o,_{\rho _{1}}}\} \\
\\
\mathbf{a}_{2}=\hbox {grad}\,\{D\,\Delta \rho _{1}+C_{2}\,\Delta \rho
_{2}-g\,_{o,_{\rho _{2}}}\}\, .
\end{array}%
\right.   \label{motions 2}
\end{equation}%
Taking Eq. (\ref{mass2}) into account, we can note that the two equations of
system (\ref{motions 1}) are equivalent to the system%
\begin{equation*}
\frac{\mathbf{\partial }\rho _{i}\mathbf{v}_{i}}{\partial t}+\func{div}%
\left( \rho _{i}\mathbf{v}_{i}\otimes \mathbf{v}_{i}\right) +\func{grad}%
\left( \rho _{i}\,\mu _{i}\right) =\mu _{i}\func{grad}\rho _{i},\ \ \
i=\{1,\ 2\}  \label{non divergence form}
\end{equation*}%
for which equations of components are not in divergence form, but the
summation of the two equations and the fact that $\ \sum_{i=1}^{2}\mu _{i}%
\func{grad}\rho _{i}=\func{grad} e\ $ allow to obtain the total motion of
the mixture in divergence form \cite{Gouin4,Gouin-Ruggeri}.
Therefore, while the global momentum equation   represents a balance law, individually, the equations of system (\ref{non divergence form}) do not.
\section{Travelling waves for an isothermal fluid mixture in one-dimensional
case}

\subsection{The system of equations for travelling waves}

An interface in a two-phase mixture is generally schematized by a surface
without thickness. Far from critical conditions, this layer is of nanometre
size and density and entropy gradients are very large. This is not the case
in the immediate vicinity of critical points where fluctuations of densities
strongly diverge. Our goal is to schematize the average of these
fluctuations by means of travelling waves in a continuous model using an
energy in form (\ref{nonlocal energy}). \newline
To obtain one-dimensional travelling waves in isothermal case when we
neglect diffusion and viscosity, we are looking for solutions only depending
on the variable \cite{Boillat,Gouin6,Hadamard,whitham}
\begin{equation*}
\xi =x-st,
\end{equation*}%
where $s$ denotes the wave velocity. Equations  (\ref{mass2}) of
conservation of masses write
\begin{equation*}
{\frac{\partial \rho _{i}}{\partial t}}+\frac{\partial (\rho _{i}{v}_{i})}{%
\partial x}=0,
\end{equation*}
where $v_i$ denotes the one-dimensional velocity of component $i$.
Equivalently  \cite{Slemrod1},
\begin{equation}                       \frac{d}{d\xi }\{\rho _{i}(v_{i}-s)\}=0\qquad \Longleftrightarrow \qquad v_{i}=%
\frac{d_{i}}{\rho _{i}}+s,  \label{balancemasses}
\end{equation}%
where $d_{i},\ i=\{1,2\}$, are constant along the motion and have the
dimension of a mass flow. The acceleration of mixture component \emph{i} is
\begin{equation*}
a_{i}={\frac{\partial v_{i}}{\partial x}}\,v_{i}+\frac{\partial v_{i}}{%
\partial t}\quad \Longleftrightarrow \quad a_{i}=\frac{d}{d\xi }\left\{
\frac{(v_{i}-s)^{2}}{2}\right\} \qquad \Longleftrightarrow \qquad a_{i}={%
\frac{1}{2}}\,{\frac{d}{dx}}\,\left({\frac{d_{i}^{2}}{\rho _{i}^{2}}}\right)
\end{equation*}%
and system (\ref{motions 2}) yields
\begin{equation}
\left\{
\begin{array}{l}
\displaystyle C_{1}\,\rho _{1}^{\prime \prime }+D\,\rho _{2}^{\prime \prime
}=\mu _{1}+{\frac{1}{2}}{\frac{d_{1}^{2}}{\rho _{1}^{2}}}+k_{1} \\
\\
\displaystyle D\,\rho _{1}^{\prime \prime }+C_{2}\,\rho _{2}^{\prime \prime
}=\mu _{2}+{\frac{1}{2}}{\frac{d_{2}^{2}}{\rho _{2}^{2}}}+k_{2}%
\end{array}%
\right. \quad \text{ or }\quad \mathbf{RX}^{\prime \prime }=\left( \frac{%
\partial Y}{\partial \mathbf{X}}\right) ^{T},   \label{dynamical system}
\end{equation}%
where $k_{1}$ and $k_{2}$ are two constants, superscript $^{T}$ denotes the
transposition and
\begin{equation*}
\mathbf{X}=\left(
\begin{array}{l}
\rho _{1} \\
\rho _{2}%
\end{array}%
\right) ,\text{ }\mathbf{R}=\left(
\begin{array}{ll}
C_{1} & D \\
D & C_{2}%
\end{array}%
\right) ,\ Y(\rho _{1},\rho _{2})=g_{o}(\rho _{1},\rho _{2})-{\frac{1}{2}}\,{%
\frac{d_{1}^{2}}{\rho _{1}}}-{\frac{1}{2}}\,{\frac{d_{2}^{2}}{\rho _{2}}}%
+k_{1}\rho _{1}+k_{2}\rho _{2}.
\end{equation*}%
Here $\ \displaystyle^{\prime }= \displaystyle {d}/{d\xi }$ denotes the derivation along the ray. Let us note that the sum of equations of
system ({\ref{dynamical system}), respectively multiplied by }$\rho
_{1}^{\prime }$ and $\rho _{2}^{\prime }$, {yields the first integral}
{\begin{equation*}
g_{o}(\rho _{1},\rho _{2})+k_{1}\,\rho _{1}+k_{2}\,\rho _{2}-{\frac{1}{2}}\,{
\frac{d_{1}^{2}}{\rho _{1}}}-{\frac{1}{2}}\,{\frac{d_{2}^{2}}{\rho _{2}}}\,- \left(
{\frac{1}{2}}\,C_{1}\,\rho _{1}^{^{\prime }2}+D\,\rho _{1}^{^{\prime
}}\,\rho _{2}^{^{\prime }}+{\frac{1}{2}}\,C_{2}\,\rho _{1}^{^{\prime
}2} \right)=k_{3},
\end{equation*}%
where $k_{3}$ is a constant.\newline
Consequently, in one-dimensional system, the profiles of mixture densities
are associated with a \emph{mechanical system} having an equivalent \emph{%
kinetic energy}
\begin{equation*}
{\frac{1}{2}}\,C_{1}\,\rho _{1}^{^{\prime }2}+D\,\rho _{1}^{^{\prime
}}\,\rho _{2}^{^{\prime }}+{\frac{1}{2}}\,C_{2}\,\rho _{1}^{^{\prime }2}
\end{equation*}%
and an equivalent \emph{potential of forces}
\begin{equation*}
-g_{o}(\rho _{1},\rho _{2})-k_{1}\,\rho _{1}-k_{2}\,\rho _{2}+{\frac{1}{2}}\,%
{\frac{d_{1}^{2}}{\rho _{1}}}+{\frac{1}{2}}\,{\frac{d_{2}^{2}}{\rho _{2}}}\,\, .
\end{equation*}%
It is possible to use all the tools of analytical dynamics as Lagrange or
Hamilton equations, symplectic geometry or Maupertuis methods. }\newline

In each bulk, density gradients are null. Eliminating constants $%
k_{i},i=\{1,2\}$, dynamical conditions through the interfacial layer yield
\begin{equation}
\left\{
\begin{array}{l}
  \left[ g_{,\rho _{1}}(\rho _{1},\rho _{2})\right]   =0 \\
  \left[ g_{,\rho _{2}}(\rho _{1},\rho _{2})\right]   =0\  \\
  \left[ g-\rho _{1}\,g_{,\rho _{1}}-\rho _{2}\,g_{,\rho _{2}}\right]
  =0 \,\, ,
\end{array}%
\right.  \label{dynamic system}
\end{equation}

where $  \left[ {u }\right]  = u_\beta-u_\alpha $ denotes the difference of values of $u$
between the two mixture bulks $\alpha$ and $\beta$ and
\begin{equation*}
g=g_{o}-{\frac{1}{2}}\,{\frac{d_{1}^{2}}{\rho _{1}}}-{\frac{1}{2}}\,{\frac{%
d_{2}^{2}}{\rho _{2}}}\,\, .
\end{equation*}%
In case of equilibrium at a given temperature $T$, ($d_{1}$ and $d_{2}$ are
null), the minimum of the total free energy, with a given total mass for
each constituent, yields conditions
\begin{equation}
\left\{
\begin{array}{l}
 \left[ \mu _{1}(\rho _{1},\rho _{2})\right]   =0 \\
  \left[ \mu _{2}(\rho _{1},\rho _{2})\right]  =0\  \\
  \left[ P(\rho _{1},\rho _{2})\right]   =0 \,\,  ,
\end{array}%
\right.  \label{static bulks}
\end{equation}%
where
\begin{equation*}
P(\rho _{1},\rho _{2})= P_c + \rho _{1}\mu _{1}(\rho _{1},\rho _{2})+\rho
_{2}\mu _{2}(\rho _{1},\rho _{2})-g_{o}(\rho _{1},\rho _{2})
\end{equation*}%
denotes the total thermodynamical pressure of the binary mixture.\ The
constant $P_c$, corresponding to the fact that $g_{o}$ is defined to an
additive constant, can be defined as the critical pressure term for a value
of $g_o$ null at the critical point. \newline
Conditions of system (\ref{dynamic system}) (or system (\ref{static bulks}))
express that the bulks of Gibbs surface $(\Sigma )$ associated with the
dynamical volume free energy $g$ (or static volume free energy $g_{o}$)
correspond to the contact points with a bitangent plane (Fig. 2). By adding
the term $\displaystyle-{\frac{1}{2}}\,{\frac{d_{1}^{2}}{\rho _{1}}}-{\frac{1%
}{2}}\,{\frac{d_{2}^{2}}{\rho _{2}}}$ to $g_{o}$, the study of
non-dissipative travelling waves of density components of a mixture turns
back to an equivalent equilibrium problem.
\begin{figure}[h]
\begin{center}
\includegraphics[width=17cm]{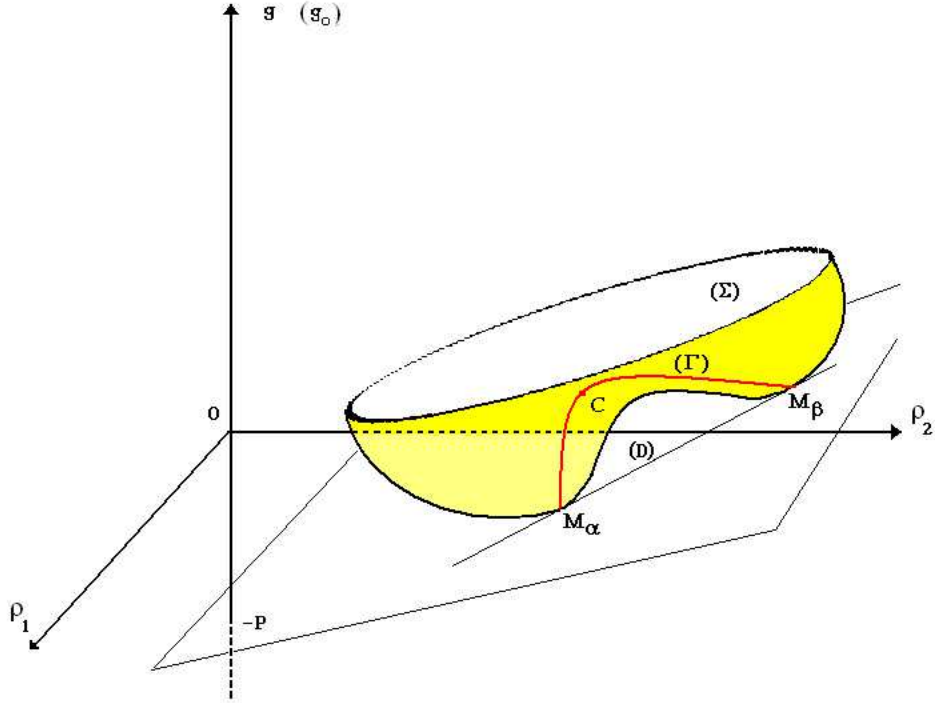}
\end{center}
\caption{In case of bulk equilibrium, the third coordinate of the point of
intersection of the bi-tangent plane with the $g$-axis is $\;-P(\protect\rho %
_{1},\protect\rho _{2})$. In dynamical case, the intersection corresponds to
the opposite of a \emph{dynamical pressure} (and defined to an additive
constant $P_c$). Contact points ${M}_\protect\alpha$ and ${M}_\protect\beta$
of the bitangent plane with surface $(\Sigma )$ generate a  curve $%
(\Gamma )$. Along this \emph{spinodal} curve $(\Gamma )$, the straight line (%
$D$) connecting the contact points ${M}_\protect\alpha$ and ${M}_\protect%
\beta$ associated with the two mixture bulks is the \emph{characteristic line%
} of the plane which generates a developable surface. Critical point $C$
corresponds to the collapse of the two contact points at the given
temperature $T$.}
\label{Fig. 2}
\end{figure}

\section{Integration of the equations of travelling waves}

In Appendix B, we propose an expansion form of the volume energy $g_{o}$ for
the binary mixture, near its critical point at temperature $T$, in the form
\begin{equation*}
g_{o}(q_{1},q_{2})=A\left\{ \left( B^{2}q_{1}^{2}-q_{2}\right) ^{2}+q_{2}^{2}%
\right\} ,
\end{equation*}%
where $A$ and $B$ are physical constants, $q_{1}$ and $q_{2}$\ are
convenient linear combinations of $\rho _{1}-\rho _{1}^{c}$ and $\rho
_{2}-\rho _{2}^{c}$ with $\rho _{1}^{c}$\ and $\rho _{2}^{c}$ denoting the
density values of the two constituents at the critical point (for the sake
of simplicity, we denote the volume free energy as $g_{o} $ in all systems
of variables). With the change of variables,
\begin{equation}
\mathbf{{r}=\left(
\begin{array}{l}
r_{1} \\
r_{2}%
\end{array}%
\right) \equiv \left(
\begin{array}{l}
\rho _{1}-\rho _{1}^{c} \\
\rho _{2}-\rho _{2}^{c}%
\end{array}%
\right) }  \label{Change 1}
\end{equation}
and
\begin{equation*}
Y(r_{1},r_{2})\equiv g_{o}(r_{1},r_{2})-{\frac{1}{2}}\,{\frac{d_{1}^{2}}{%
r_{1}+\rho _{1}^{c}}}-{\frac{1}{2}}\,{\frac{d_{2}^{2}}{r_{2}+\rho _{2}^{c}}}%
+k_{1}\left( r_{1}+\rho _{1}^{c}\right) +k_{2}\left( r_{2}+\rho
_{2}^{c}\right)
\end{equation*}%
system (\ref{dynamical system}) can be written%
\begin{equation}
\mathbf{R}\ \mathbf{r}^{\prime \prime }=\left( \frac{\partial Y}{\partial
\mathbf{r}}\right) ^{T}.  \label{NdynaSyst}
\end{equation}
We denote
\begin{equation*}
\mathbf{{P}=\left(
\begin{array}{ll}
a\  & \ b \\
c\  & \ d%
\end{array}%
\right) ,}  \label{matrix of change}
\end{equation*}%
where scalars $a, b, c, d$ \, are depending on the thermo-mechanical
properties of the mixture near the point $C$. By the new change of
variables,
\begin{equation}
\mathbf{{r}={P\,q} \,\, ,\text{\ with\ \ }{q}=\left(
\begin{array}{l}
q_{1} \\
q_{2}%
\end{array}%
\right)}  \label{Change 2}
\end{equation}
and
\begin{eqnarray*}
Y(q_{1},q_{2}) &=&g_{o}(q_{1},q_{2})-{\frac{1}{2}}\,{\frac{d_{1}^{2}}{%
aq_{1}+bq_{2}+\rho _{1}^{c}}}-{\frac{1}{2}}\,{\frac{d_{2}^{2}}{%
cq_{1}+dq_{2}+\rho _{2}^{c}}} \\
&&+k_{1}\left( aq_{1}+bq_{2}+\rho _{1}^{c}\right) +k_{2}\left(
cq_{1}+dq_{2}+\rho _{2}^{c}\right)
\end{eqnarray*}%
the system (\ref{NdynaSyst}) writes in the form
\begin{equation*}
\widetilde{\mathbf{R}}\ \mathbf{q}^{\prime \prime }=\left( \frac{\partial Y}{%
\partial \mathbf{q}}\right) ^{T}\text{ \ \ \ \ \ \ with\ \ \ \ \ \ \ }%
\widetilde{\mathbf{R}}=\left(
\begin{array}{ll}
\widetilde{C}_{1} & \ \widetilde{D} \\
\widetilde{D}\  & \ \widetilde{C}_{2}%
\end{array}%
\right)  \label{final dyna system}
\end{equation*}%
or
\begin{equation}
\left\{
\begin{array}{l}
\widetilde{C}_{1}q_{1}^{\prime \prime }+\widetilde{D}\ q_{2}^{\prime \prime
}=4AB^{2}\left( B^{2}q_{1}^{3}-q_{1}q_{2}\right) +\widetilde{k}_{1} \\
\widetilde{D}\ q_{1}^{\prime \prime }+\widetilde{C_{2}}q_{2}^{\prime \prime
}=2A\left( -B^{2}q_{1}^{2}+2q_{2}\right) +\widetilde{k}_{2} \,\, ,
\end{array}%
\right.  \label{final system develo}
\end{equation}%
where $\widetilde{k}_{1}=ak_{1}+ck_{2}$ and $\widetilde{k}_{2}=bk_{1}+dk_{2}$. Due to the
fact $\mathbf{R}$ is symmetric positive definite  and $\mathbf P$ is not singular,   the
same holds for $\widetilde{\mathbf{R}}=\mathbf{P}^{T}\mathbf{R\,P}$.

\subsection{Case of equilibrium}

\subsubsection{The rescaling process}

The values of constants $\widetilde{k}_{1}$ and $\widetilde{k}_{2}$ must
correspond to the singular points of the differential system (\ref{dynamical
system}) when $d_{i}=0,\ (i=1,2)$. These singular points are associated with
the two bulks of the binary mixture. Consequently, $\widetilde{k}%
_{1}=k_{1}=0 $ \ and $\widetilde{k}_{2}=k_{2}=-\tau ^{2}$ (see Appendix B,
for the meaning of $\tau $ as a \emph{"distance"} to the critical point).

We consider the following hypothesis $(H_{1})$ and $(H_{2})$:

$\bullet \ {(H_{1})}$\quad We assume that $q_{1}$ and $q_{2}$ slowly vary as
a function of the one-dimensional parameter $x$. The hypothesis can be
expressed by a change of variable $z=\varepsilon \ x$, where $\varepsilon $
is an adimensional small parameter ($\varepsilon \ll \ 1$). With the
following change of variables:
\begin{equation*}
q_{1}(x)=Q_{1}(z)\text{ \ \ and \ \ }q_{2}(x)=Q_{2}(z)\,,
\end{equation*}%
system (\ref{final system develo}) yields
\begin{equation}
\left\{
\begin{array}{l}
\varepsilon ^{2}\left( \widetilde{C}_{1}\ddot{Q}_{1}+\widetilde{D}{\ }\ddot{Q%
}_{2}\right) =4AB^{2}\left( B^{2}Q_{1}^{3}-Q_{1}Q_{2}\right) , \\
\varepsilon ^{2}\left( \widetilde{D}{\ }\ddot{Q}_{1}+\widetilde{C}_{2}\ddot{Q%
}_{2}\right) =2A\left( -B^{2}Q_{1}^{2}+2Q_{2}\right) -\tau ^{2},%
\end{array}%
\right.  \label{premiere rescale}
\end{equation}%
where the dot denotes the derivation with respect to $z$. The
solution must be a phase transition connecting singular points $M_{\alpha
}(\tau /B,\tau ^{2})$ and $M_{\alpha }(-\tau /B,\tau ^{2})$.

$\bullet \ {(H_{2})}$\quad The density profiles are of small amplitude with
respect to the critical densities. Consequently, we look for the solutions
in the form
\begin{equation*}
Q_{1}(z)=\varepsilon ^{n_{1}}y_{1}(z)\quad \text{and}\quad
Q_{2}(z)=\varepsilon ^{n_{2}}y_{2}(z)\, ,
\end{equation*}%
where $n_{1}$ and $n_{2}$ are to real positive constants. System (\ref%
{premiere rescale}) yields
\begin{equation}
\left\{
\begin{array}{l}
\varepsilon ^{2}\left( \widetilde{C}_{1}\varepsilon ^{n_{1}}\ddot{y}_{1}+%
\widetilde{D}{\ }\varepsilon ^{n_{2}}\ddot{y}_{2}\right) =4AB^{2}\left(
B^{2}\varepsilon ^{3n_{1}}y_{1}^{3}-\varepsilon
^{n_{1}+n_{2}}y_{1}y_{2}\right) , \\
\varepsilon ^{2}\left( \widetilde{D}{\ }\varepsilon ^{n_{1}}\ddot{y}_{1}+%
\widetilde{C}_{2}\varepsilon ^{n_{2}}\ddot{y}_{2}\right) =2A\left(
-B^{2}\varepsilon ^{2n_{1}}y_{1}^{2}+2\varepsilon ^{n_{2}}y_{2}\right) -\tau
^{2}.%
\end{array}%
\right.  \label{deuxieme rescale}
\end{equation}%
The form of the volume energy near the critical point must be conserved by
affinity. This assumption corresponds to the universality of the form of the
energy near a critical point \cite{Domb}. Consequently in this rescaling
process,
\begin{equation*}
2n_{1}=n_{2}\text{ \ \ and \ \ \ }\tau =\varepsilon ^{n_{1}}E\, ,
\end{equation*}%
where $E$ is a positive constant with physical dimension $M^{1/2}L^{-3/2}$. System (\ref{deuxieme rescale}) yields the equality of same
order terms,
\begin{equation}
\left\{
\begin{array}{l}
\varepsilon ^{2+n_{1}}\widetilde{C}_{1}\ddot{y}_{1}=\varepsilon
^{3n_{1}}4AB^{2}\left( B^{2}y_{1}^{3}-y_{1}y_{2}\right) , \\
\varepsilon ^{2+n_{1}\ }\widetilde{D}{\ }\ddot{y}_{1}=\varepsilon
^{2n_{1}}2A\left( -B^{2}y_{1}^{2}+2\ y_{2}-E^{2}\right) .%
\end{array}%
\right.  \label{troisieme rescale}
\end{equation}
If we assume $2+n_1<2\,n_1$, Eq. (\ref{troisieme rescale})$_2$ reduces to $\ddot{y}_{1}=0$, which is not
physically possible; If we assume $2+n_1=2\,n_1$, Eq. (\ref{troisieme rescale})$_1$ reduces to $\ddot{y}_{1}=0$. Then,
to obtain non-trivial solutions, we must assume that $2+n_1>2\,n_1$ and $%
2A\left( -B^{2}y_{1}^{2}+2\ y_{2}-E^{2}\right) =0$; consequently,
 \begin{equation*}
 y_{2}= \frac{1}{2}\left(B^{2}y_{1}^{2}+ E^{2}\right).
 \end{equation*}
 Now, if we compare $\varepsilon ^{2+n_{1}}$ and $\varepsilon ^{3n_{1}}$, we must assume that $n_1=1$.
 System (\ref{troisieme
rescale}) leads to the \emph{rescaled system}
\begin{equation*}
\left\{
\begin{array}{l}
\widetilde{{C}}_{1}\ddot{y}_{1}=2AB^{2}\left(
B^{2}y_{1}^{3}-E^{2}y_{1}\right)  \\
\displaystyle{\ }y_{2}=\frac{1}{2}\left( B^{2}y_{1}^{2}+E^{2} \right) %
\end{array}%
\right.
\end{equation*}%
which is equivalent to
\begin{equation}
\left\{
\begin{array}{l}
\displaystyle\frac{\widetilde{C}_{1}}{2}\,\dot{y}_{1}^{2}=AB^{2}\left( \frac{%
B^{2}}{2}y_{1}^{4}-E^{2}y_{1}^{2}+k_{4}\right)  \\
\displaystyle{\ }y_{2}=\frac{1}{2}\left( B^{2}y_{1}^{2}+E^{2}\right) ,%
\end{array}%
\right.  \label{integrate recaled sytem}
\end{equation}%
where $k_{4}$ is a convenient constant of integration.

\subsubsection{Integration of system (\protect\ref{integrate recaled sytem})}

The first equation of system (\ref{integrate recaled sytem}) has several
possible solutions. In this case we look for phase transition, the solution
must be monotonic and must go from an equilibrium point to the other one. The
simplest possibility is $k_{4}=E^{4}/(2B^{2})$. The integration of Eq. (\ref%
{integrate recaled sytem})$_{1}$ yields%
\begin{equation*}
y_{1}(z)=\pm \frac{E}{B}\tanh \left( E\sqrt{\frac{2AB^{2}}{\widetilde{C}_{1}}%
}x\right)\,.
\end{equation*}%
 Consequently, with the change
of orientation of the $x$-axis, we get
\begin{equation*}
\left\{
\begin{array}{l}
\displaystyle q_{1}(x)=\frac{\tau}{B}\tanh \left( \tau\sqrt{\frac{2AB^{2}}{%
\widetilde{C}_{1}}}x\right)  \\
\displaystyle q_{2}(x)=\frac{\tau^{2}}{2}\left\{ 1+\tanh ^{2}\left( \tau\sqrt{\frac{%
2AB^{2}}{\widetilde{C}_{1}}}x\right) \right\} .%
\end{array}%
\right.
\end{equation*}%
Due to the changes of variables (\ref{Change 1}), (\ref{Change 2}), we
obtain the general form of the phase transitions
\begin{equation}
\left\{
\begin{array}{l}
\displaystyle\rho _{1}(x)=\rho _{1}^{c}+a_{1}\tanh \left( \frac{x}{L}\right)
+b_{1}\left\{ 1+\tanh ^{2}\left( \frac{x}{L}\right) \right\}  \\
\\
\displaystyle\rho _{2}(x)=\rho _{2}^{c}+a_{2}\tanh \left( \frac{x}{L}\right)
+b_{2}\left\{ 1+\tanh ^{2}\left( \frac{x}{L}\right) \right\} .%
\end{array}%
\right.  \label{Sol R}
\end{equation}%
Depending on the two cases of boundary conditions, we obtain

1$^{st}$ Case: When $\displaystyle\lim_{x\rightarrow -\infty }\rho
_{i}(x)=\rho _{i}^{\alpha }$ \ and \ $\displaystyle\lim_{x\rightarrow
+\infty }\rho _{i}(x)=\rho _{i}^{\beta }$ \ with $i=1,2$ , then
\begin{equation*}
a_{i}=\frac{1}{2}\left( \rho _{i}^{\beta }-\rho _{i}^{\alpha }\right), \quad
b_{i}=\frac{1}{4}\left( \rho _{i}^{\alpha }-2\rho _{i}^{c}+\rho _{i}^{\beta
}\right).
\end{equation*}
2$^{nd}$ Case: When $\displaystyle\lim_{x\rightarrow -\infty }\rho
_{i}(x)=\rho _{i}^{\beta }$ \ and \ $\displaystyle\lim_{x\rightarrow +\infty
}\rho _{i}(x)=\rho _{i}^{\alpha }$ \ with $i=1,2$ , then
\begin{equation*}
a_{i}=\frac{1}{2}\left( \rho _{i}^{\alpha }-\rho _{i}^{\beta }\right), \quad
b_{i}=\frac{1}{4}\left( \rho _{i}^{\alpha }-2\rho _{i}^{c}+\rho _{i}^{\beta
}\right).
\end{equation*}
Let us note that the constants $a_i$ and $b_i$ depend on the thermodynamical
behavior of the mixture and they may be positive or negative.

\subsubsection{The different cases}

The variation of $\rho _{1}(x)$ and $\rho _{2}(x)$ in system (\ref{Sol R})
yields the different cases. They are summarized on Fig. 3 as domains
delimited by the bisectrices of axes $a_{i}$ and $b_{i}$ and we represent
the different possible variations of each component on Fig. 4.
\begin{figure}[h]
\begin{center}
\includegraphics[width=7cm]{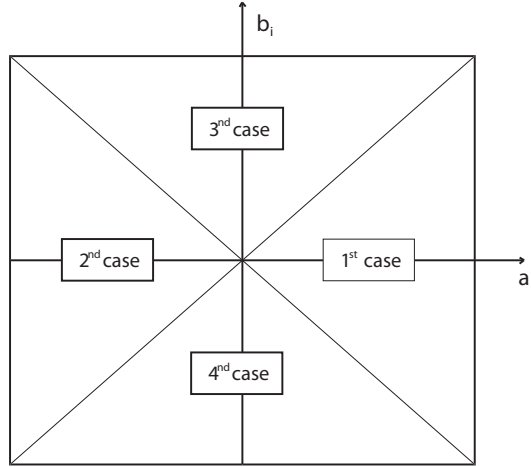}
\end{center}
\caption{The different possible cases for each mixture component are
presented on the graph associated with the $a_{i}$-axis and $b_{i}$-axis.
Points on axis $a_i$ where $\protect\rho_i^\protect\alpha=\protect\rho_i^%
\protect\beta$ are forbidden; this will not be the case in dynamics.}
\label{Fig. 3}
\end{figure}
\begin{figure}[h]
\begin{center}
\includegraphics[width=9cm]{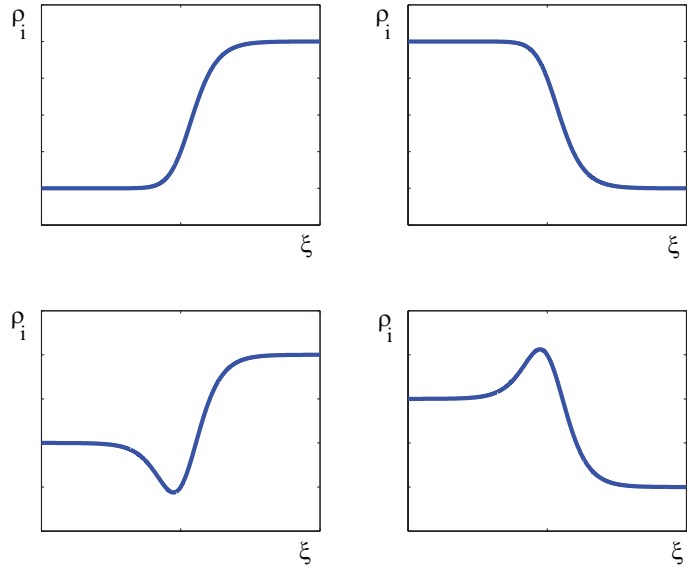}
\end{center}
\caption{Sketch of the different possible forms of phase transition for each mixture
component are presented on the graph with densities $\protect\rho%
_{i}$ versus  $\xi$.}
\label{Fig. 4}
\end{figure}
We see, as it is known by other methods in physical chemistry  that the
density profiles in interfacial densities are not necessary homogeneous for
mixtures of fluids \cite{Rowlinson1,Rowlinson2}.

\subsection{Case of travelling waves}

In  the case of motion, we check with travelling waves submitted to viscosity;   in this case,   system (\ref{motions 2}) can
be modified to take account of the viscosity in each component of the
mixture
\begin{equation*}
\left\{
\begin{array}{l}
\mathbf{a}_{1}=\displaystyle\hbox {grad}\,\{C_{1}\,\Delta \rho _{1}+D\,\Delta \rho
_{2}-g\,_{o,_{\rho _{1}}}\}+\frac{1}{\rho _{1}}\func{div}\mathbf{\sigma }_{1}
\\
\\
\mathbf{a}_{2}=\displaystyle\hbox {grad}\,\{D\,\Delta \rho _{1}+C_{2}\,\Delta \rho
_{2}-g\,_{o,_{\rho _{2}}}\}+\frac{1}{\rho _{2}}\func{div}\mathbf{\sigma }_{2}\, \, ,
\end{array}%
\right.  \label{motions 3}
\end{equation*}%
where $\mathbf{\sigma }_{i} \left\{ i=1,2\right\} $ is the viscous stress
tensor associated with component $i$. Consequently we obtain in variables $%
q_{1}$ and $q_{2}$, the system of equations of motions in one-dimensional
system with $\sigma _{i}=\nu _{i}\,v_{i}^{\prime \prime }$, where $\nu _{i}>0$ is the
coefficient of viscosity of the component $i$ \cite{Sampaio}.\ The system can be
written by using one dimensional derivation in $\func{div}\mathbf{\sigma }%
_{i},$ by integration as in section 4.1 and with the fact that $\rho
_{i}\approx \rho _{i}^{c}$,
\begin{equation}
\left\{
\begin{array}{l}
\displaystyle\widetilde{C}_{1}q_{1}^{\prime \prime }+\widetilde{D}\
q_{2}^{\prime \prime }=4AB^{2}\left( B^{2}q_{1}^{3}-q_{1}q_{2}\right) \\
\displaystyle+\frac{1}{2}\frac{a\ d_{1}^{2}}{\left( \rho _{1}^{c}\right) ^{2}%
}+\frac{1}{2}\frac{c\ d_{2}^{2}}{\left( \rho _{2}^{c}\right) ^{2}}-\frac{%
a\,d_{1}\,\nu _{1}}{\left( \rho _{1}^{c}\right) ^{3}}\,q_{1}^{\prime }-\frac{%
c\,d_{2}\,\nu _{2}}{\left( \rho _{1}^{c}\right) ^{3}}\,q_{2}^{\prime }+%
\widetilde{k}_{1} \\
\displaystyle\widetilde{D}\ q_{1}^{\prime \prime }+\widetilde{C_{2}}%
q_{2}^{\prime \prime }=2A\left( -B^{2}q_{1}^{2}+2q_{2}\right) \\
\displaystyle+\frac{1}{2}\frac{b\ d_{1}^{2}}{\left( \rho _{1}^{c}\right) ^{2}%
}+\frac{1}{2}\frac{d\ d_{2}^{2}}{\left( \rho _{2}^{c}\right) ^{2}}-\frac{%
b\,d_{1}\,\nu _{1}}{\left( \rho _{1}^{c}\right) ^{3}}\,q_{1}^{\prime }-\frac{%
d\,d_{2}\,\nu _{2}}{\left( \rho _{1}^{c}\right) ^{3}}\,q_{2}^{\prime }+%
\widetilde{k}_{2}\, .
\end{array}%
\right.  \label{system travelling waves}
\end{equation}

\subsubsection{The rescaling process}

We can make an analogous rescaling process as in Section 4.1.1.\ The
difference only comes from scalars $d_{1}$ and $d_{2}$. They must be of
small amplitude; consequently, we look for constants in the form
\begin{equation*}
D_{1}=\varepsilon ^{n_{3}}d_{1} \quad \text{and}\quad D_{2}=\varepsilon
^{n_{4}}d_{2},
\end{equation*}%
where constants $n_{3}$ and $n_{4}$ are positive. To obtain convenient
solutions we must have
\begin{equation*}
d_{i}=D_{i}\varepsilon ^{\frac{3n_{1}}{2}},(i=1,2),\ \widetilde{k}%
_{1}=\varepsilon ^{3n_{1}}\widetilde{K}_{1},\widetilde{k}_{2}=\varepsilon
^{2n_{1}}\ \widetilde{K}_{2},\ n_{1}=1.
\end{equation*}%
Then, the  terms $\displaystyle\frac{d_{1}\,\nu _{1}}{\left( \rho _{1}^{c}\right)
^{3}}\,q_{1}^{\prime }=\frac{D_{1}\,\nu _{1}}{\left( \rho _{1}^{c}\right)
^{3}}\,\varepsilon ^{\left( 3+\frac{_{1}}{2}\right) }\dot{y}_{1}$ and $%
\displaystyle\frac{d_{2}\,\nu _{2}}{\left( \rho _{2}^{c}\right) ^{3}}%
\,q_{2}^{\prime }=\frac{D_{2}\,\nu _{2}}{\left( \rho _{2}^{c}\right) ^{3}}%
\,\varepsilon ^{\left( 4+\frac{_{1}}{2}\right) }\dot{y}_{2}$ which are of
order greater than the other terms in system (\ref{system travelling waves}%
), are negligible. Finally, system (\ref{system travelling waves}) yields
\begin{equation}
\left\{
\begin{array}{l}
\displaystyle\widetilde{{C}}_{1}\dot{y}_{1}^{2}=AB^{2}\left( B^{2}y_{1}^{4}+%
\frac{\widetilde{K}_{2}}{A}y_{1}^{2}+\widetilde{K}_{3}y_{1}+\widetilde{K}%
_{4}\right) , \\
\displaystyle{\ }y_{2}=\frac{1}{2}\left( B^{2}y_{1}^{2}-\frac{\widetilde{K}%
_{2}}{2A}\right) ,%
\end{array}%
\right.  \label{trav integrated}
\end{equation}%
where $\displaystyle\widetilde{K}_{3}=\frac{2}{AB^{2}}\left( \frac{a\
D_{1}^{2}}{2\ \left( \rho _{1}^{c}\right) ^{2}}+\frac{c\ D_{2}^{2}}{2\
\left( \rho _{2}^{c}\right) ^{2}}+\widetilde{K}_{1}\right) $ and $\widetilde{%
K}_{4}$\ is a new constant of integration.

\subsubsection{Integration of system (\protect\ref{trav integrated})}

A qualitative study of  Eq. (\ref{trav integrated})$_{1}$ can be
considered. Constant $A$ is strictly positive. The discussion comes from
\begin{equation*}
\dot{y}_{1}^{2}=f(y_{1})-g(y_{1}),
\end{equation*}
where
\begin{equation}
f(y_{1})=\frac{AB^{4}}{\widetilde{{C}}_{1}}\left( y_{1}^{2}+\frac{\widetilde{%
K}_{2}}{2AB^{2}}\right) ^{2}  \label{quartic}
\end{equation}
represents a quartic and
\begin{equation}
g(y_{1})=\frac{AB^{2}}{\widetilde{{C}}_{1}}\left( -\widetilde{K}_{3}y_{1}-%
\widetilde{K}_{4}+\frac{\widetilde{K_{2}}^{2}}{4A^{2}B^{2}}\right)
\label{line}
\end{equation}
represents a straight line. If $\widetilde{K}_{2}\geq 0,\ f(y_{1})$ is a convex
quartic and only periodic solutions are possible. Consequently $\widetilde{K}%
_{2}<0$.

Two interesting cases can be considered:

a) The straight line (\ref{line}) is bitangent to the quartic (\ref{quartic}). The
solution of Eq. (\ref{trav integrated})$_{1}$ is a travelling wave of phase
transition in the same form than the two first curves of Fig.\ 4. We must
have
\begin{equation*}
\widetilde{K}_{3}=0,\quad \widetilde{K}_{4}=\frac{\widetilde{K_{2}}^{2}}{%
4A^{2}B^{2}}\quad \text{ \ and consequently }\quad \widetilde{K}_{1}=-\frac{%
aD_{1}^{2}}{2\left( \rho _{1}^{c}\right) ^{2}}\text{\ }-\frac{cD_{2}^{2}}{%
2\left( \rho _{2}^{c}\right) ^{2}}\,\, \text{.}
\end{equation*}
Then,%
\begin{equation*}
\left\{
\begin{array}{l}
\displaystyle q_{1}(\xi )=\frac{\varkappa }{B}\tanh \left( \varkappa \sqrt{%
\frac{2AB^{2}}{\widetilde{{C}}_{1}}}\xi \right) \\
\\
\displaystyle q_{2}(\xi )=\frac{\varkappa ^{2}}{2}\left\{ 1+\tanh ^{2}\left(
\varkappa \sqrt{\frac{2AB^{2}}{\widetilde{{C}}_{1}}}\xi \right) \right\}  .
\end{array}%
\right. \text{ \ \ with \ \ }\varkappa =\varepsilon \sqrt{\frac{-\widetilde{%
K_{2}}}{2A}}
\end{equation*}
We note that $\varkappa $ plays the role of $\tau$ in the static case. The
solutions in variables $\rho _{1}$ and $\rho _{2}$ are in the same form than
in Section 4.1.2 and the same sort of density profiles can be observed
\begin{equation*}
\left\{
\begin{array}{l}
\displaystyle\rho _{1}(\xi )=\rho _{1}^{c}+\frac{a\varkappa }{B}\tanh \left(
\frac{\xi }{L_{1}}\right) +\frac{b\varkappa ^{2}}{2}\ \left\{ 1+\tanh
^{2}\left( \frac{\xi }{L_{1}}\right) \right\}  \\
\\
\displaystyle\rho _{2}(\xi )=\rho _{2}^{c}+\frac{c\varkappa }{B}\tanh \left(
\frac{\xi }{L_{1}}\right) +\frac{d\varkappa ^{2}}{2}\left\{ 1+\tanh
^{2}\left( \frac{\xi }{L_{1}}\right) \right\} .%
\end{array}%
\right. \text{with }L_{1}=\frac{1}{\varkappa }\sqrt{\frac{\widetilde{{C}}%
_{1}}{2AB^{2}}}
\end{equation*}
We have seen that to obtain travelling waves near a critical point, the viscosity terms have to be negligible with respect to the others. This result is in accordance with dispersive waves as the ones obtained by using the Korteweg-de Vries equation \cite{whitham}.

\section{Conclusion}

The equations of binary mixtures of fluids are considered in the
one-dimensional case. By using an asymptotic expansion of the volume free
energy near a critical point and by taking the density gradients into
account, we get, in continuum mechanics, a dynamical system representing the
equations of travelling waves for the two-fluid components. A rescaling
process associated with the universality of the energy form near a critical
point allows to obtain different sorts of travelling waves of density as
well as in conservative and dissipative cases. We can discuss different
possible phase transitions according to the values of phases densities of
components and the characteristic of the mixture.

Now, we consider some physical comments.
For given densities, at time $t=0$, the
velocities of the mixture components are deduced  from the mass balance
equations (\ref{balancemasses}). Then, an arbitrary constant $s$ is introduced in the velocity fields:
\begin{equation*}
v_{i}(x)=\frac{d_{i}}{\rho _{i}(x)}+s
\end{equation*}
and at any time $t$   the relation leads to
\begin{equation*}
v_{i}(x-st)=\frac{d_{i}}{\rho _{i}(x-st)}+s\,.
\end{equation*}
The initial conditions yield this arbitrary velocity $s$.

\emph{First case: interface propagation -}   By crossing the interface (\emph{interface} means the
domain where the densities drastically change), the volume changes.  This phenomenon
cannot occur in a closed tube with constant volume, but only in a tube with one  closed
end.  It is similar to a phenomenon of vaporization or condensation.%
\newline
One imagines that the other side of the tube is closed with a piston whose
displacement may be imposed.\ The component velocities at the fixed end are
zero in bulk $\beta $; this determines the constant $s$ thanks to the condition
\begin{equation*}
s=\frac{d_{1}}{\rho _{1}^{\beta }}=\frac{d_{2}}{\rho _{2}^{\beta }}\  .
\end{equation*}
Due to the fact that $\rho _{1}^{\beta }\simeq\rho _{1}^{c}$ and $\rho
_{2}^{\beta }\simeq\rho _{2}^{c}$, we get
\begin{equation}
s=\frac{d_{1}}{{\rho _{1}^{c}}}=\frac{d_{2}}{\rho _{2}^{c}}
\label{debits}
\end{equation}
and the values of $d_{1}$ and $d_{2}$ are not  independent.

\emph{Second case: solitary waves -}  The volume of the interface only moves in one
bulk phase and remains constant.  Such a wave may   move in a closed tube as
Natterer's one \cite{Bruhat}. At the ends which are assumed far from the
interface, the velocities in the bulk phase are zero and we still obtain
Eq. (\ref{debits}) such that the values of $d_{1}$ and $d_{2}$ are not independent.

\section{Appendix A: Proof of system (\protect\ref{motions 2})}

Hamilton's principle - variational form of the principle of virtual powers -
allows us to derive the equations of motions. The variations of motions
of particles are deduced from the functions
\begin{equation*}
{\mathbf{X}}_i=\mathbf{\Psi}_i({\mathbf{z}}\,;\beta_i ), \quad i\in \{1,2\}
\end{equation*}%
for which $\beta _i $ are two parameters defined in a neighborhood of zero;
they are associated with two families of virtual motions of the mixture. The
real motions correspond to $\beta_i =0$ : $\quad \mathbf{\Psi}_i({\mathbf{z}}%
\,; 0)= {\Phi}_i({\mathbf{z}}) $. \newline
Virtual material displacements associated with any variation of real motions
can be written as \cite{Gouin4}
\begin{equation*}
\delta {\mathbf{X}}_i= \left.\frac {\partial \mathbf{\Psi }_i}{\partial
\beta_i }\right\vert_{\beta_i =0}.
\end{equation*}%
Such a variation is dual with Serrin's \cite{Serrin}. This has been studied
in \cite{Gouin4} and corresponds to the natural variation of the motion in a
Lagrangian representation.

Let us denote $\mathbf{\xi }_{i}=\delta _{i}\mathbf{X}_{i}$ ; we deduce the
two relations in Lagrangian variables \cite{Casal3,Lin,Serrin}
\begin{equation*}
\begin{array}{lll}
\displaystyle\delta _{i}\mathbf{v}_{i}=-\mathbf{F}_{i}\frac{\partial \mathbf{%
\xi }_{i}}{\partial t}, &  & \displaystyle\delta _{i}\rho _{i}=\rho _{i}\,{%
\func{div}}_{oi}\,\mathbf{\xi }_{i}+\frac{\rho _{i}}{\rho _{oi}}\frac{%
\partial \rho _{oi}}{\partial \mathbf{X}_{i}}\,\mathbf{\xi }_{i}\, ,%
\end{array}%
\end{equation*}%
where $\displaystyle {\partial }/{\partial \mathbf{X}_{i}}$ is the linear
form associated with the gradient and $\func{div}_{oi}$ is the divergence
operator on $\mathcal{D}_{oi}$. Assuming that terms on the edge of $
\mathcal{W}$ are zero, we obtain
\begin{equation*}
\delta _{i}a=\int_{t_{1}}^{t_{2}}\int_{\mathcal{D}_{t}}\left\{ \left( \frac{1%
}{2} \mathbf{v}_{i}^2-e_{,\rho _{i}}-\Omega _{i}\right) \delta _{i}\rho
_{i}+\rho _{i}\mathbf{v}_{i}^{T}\delta _{i}\mathbf{v}_{i}-\left\{ \frac{%
\partial }{\partial x_{\gamma }}\left( \frac{\partial e }{\partial \rho _{{i}%
,\gamma}}\right) \right\} \ \delta _{i}\rho _{i}\right\} dvdt
\end{equation*}%
or%
\begin{equation*}
\delta _{i}a=\int_{t_{1}}^{t_{2}}\int_{\mathcal{D}_{oi}}\rho _{oi}\left\{
R_{i}\,{\func{div}}_{oi}\,\mathbf{\xi }_{i}+\frac{R_{i}}{\rho _{oi}}\frac{%
\partial \rho _{oi}}{\partial \mathbf{X}_{i}}\,\mathbf{\xi }_{i}-\mathbf{v}%
_{i}^{T}\mathbf{F}_{i}\frac{\partial \mathbf{\xi }_{i}}{\partial t}\right\}
dv_{oi}dt,
\end{equation*}%
where $R_{i}$ $=\frac{1}{2} \mathbf{v}_{i}^2-\mu _{i}-\Omega _{i}$. But,
\begin{equation*}
{\func{div}}_{oi}(\rho _{oi}R_{i}\mathbf{\xi }_{i})=\rho _{oi}R_{i}{\func{div%
}}_{oi}\,\mathbf{\xi }_{i}+R_{i}\frac{\partial \rho _{oi}}{\partial \mathbf{X%
}_{i}}\,\mathbf{\xi }_{i}+\rho _{oi}\frac{\partial R_{i}}{\partial \mathbf{X}%
_{i}}\,\mathbf{\xi }_{i}.
\end{equation*}%
So, the terms given by integration on the edge of $\mathcal{D}_{oi}$ being
zero,
\begin{equation*}
\delta _{i}a=\int_{t_{1}}^{t_{2}}\int_{\mathcal{D}_{oi}}\rho _{oi}\,\left\{-%
\frac{\partial R_{i}}{\partial \mathbf{X}_{i}}+\frac{\partial }{\partial t}(%
\mathbf{v}_{i}^{T}\mathbf{F}_{i})\right\} \ \mathbf{\xi }_{i}\,dv_{oi}dt.
\end{equation*}%
Finally, for each constituent,%
\begin{equation*}
\frac{\partial }{\partial t}(\mathbf{v}_{i}^{T }\mathbf{F}_{i})=\frac{%
\partial R_{i}}{\partial \mathbf{X}_{i}}
\end{equation*}%
and by using the relation%
\begin{equation*}
\frac{\partial }{\partial t}(\mathbf{v}_{i}^{T }\mathbf{F}_{i})=\mathbf{a}%
_{i}^{T}\mathbf{F}_{i}+\mathbf{v}_{i}^{T}\frac{\partial \mathbf{v}_{i}}{%
\partial \mathbf{X}_{i}}
\end{equation*}%
we obtain
\begin{equation*}
\mathbf{a}_{i}^{T}\mathbf{F}_{i}+\frac{1}{2}\,\frac{\partial \mathbf{v}%
_{i}^2 }{\partial \mathbf{X}_{i}}\, =\frac{\partial R_{i}}{\partial \mathbf{X%
}_{i}}\qquad \text{or}\qquad \mathbf{a}_{i}^{T }+\frac{\partial }{\partial
\mathbf{x}}(\mu _{i}+\Omega _{i})=0.
\end{equation*}%
This leads to the vectorial form (\ref{motions 1})
\begin{equation*}
\mathbf{a}_{i}+\func{grad}(\mu _{i}+\Omega _{i})=0,\ i=\{1,\ 2\}.
\end{equation*}

\section{Appendix B: The volume free energy of a binary mixture near
critical conditions}

In this Appendix, thanks to a direct method of differential geometry, we obtain  the form of the free energy of a binary mixture near a critical point. In physical chemistry, the form is presented thanks to the knowledge of the thermodynamic coordinates of the bulks when we are near a critical point \cite{Rowlinson1}.
In this new presentation, we take only  account of conditions of equilibrium of bulks when the critical point marks the limit of their coexistence.
Such a method implies general physical consequences as in one component fluid when the two variables are densities of matter and entropy but might also be extended in other topics when different equilibrium states collapse  only in one state as in finance or biology.

The volume free energy of binary mixture at temperature $T$ is taken in the
form:
\begin{equation*}
g_o=g_{o}(\rho _{1},\rho _{2})\, ,
\end{equation*}%
where $\rho _{1},\rho _{2}$ are the densities of the two components. The
following relations between the chemical potentials $\mu _{1},\mu _{2}$ of
the two constituents and $g_{o}$ hold
\begin{equation*}
\mu _{1}=\left( g_{o}\right) _{\rho _{1}}^{\prime } ,\,\,\ \mu _{2}=\left(
g_{o}\right) _{\rho _{2}}^{\prime }
\end{equation*}%
and the thermodynamical pressure is given by
\begin{equation*}
P=-g_o +\rho _{1}\mu _{1}+\rho _{2}\mu _{2}+ P_c=-g_o +\rho _{1}\left(
g_{o}\right) _{\rho _{1}}^{\prime }+\rho _{2}\left( g_{o}\right) _{\rho
_{2}}^{\prime }+ P_c.
\end{equation*}
As the function $g_{o}$ is twice differentiable, the geometrical
representation of $g_{o}$ by a surface $\Sigma $ admits a tangent plane at
any point. The normal vector to the plane has the direction coefficients (-$%
\mu _{1}$ -$\mu _{2}$, 1).

The phase equilibrium of two bulks $\alpha$, $\beta$ is described by the
conditions
\begin{equation*}
T(\alpha)=T(\beta)\,\,\, , \,\,\, \mu_{1}(\alpha)= \mu_{1}(\beta)\,\,\, ,
\,\,\, \mu_{2}(\alpha)= \mu_{2}(\beta)\,\,\, , \,\,\, P(\alpha)=P(\beta).
\end{equation*}
It is noticeable that, in this case, the third coordinate of the
intersection point of the bitangent plane with the third axis is $%
-P(\rho_{1}, \rho_{2}, T)+ P_c $ (see Fig. 2). The contact points $%
M_{\alpha} $, $M_{\beta}$ of the bitangent plane to $\Sigma$ generate two
curves $(\Gamma_{\alpha})$, $(\Gamma_{\beta})$ respectively. The curve $%
(\Gamma)= (\Gamma_{\alpha}) \cup (\Gamma_{\beta})$ is the \textit{spinodal}
curve of the phase transition. The case with only one contact point
corresponds to a mixture with only one bulk while two contact points with a
bitangent plane correspond to  a two-bulk equilibrium mixture. \newline
When, at a given temperature $T$, $M_{\alpha}$ tends to $M_{\beta}$, the
limit point $C$ is a \textit{critical point} corresponding to the fact that
two bulks collapse into one bulk. The study of the mixture equilibrium near
the critical point $C$, at a given temperature $T$, is closely related to
the shape of surface $\Sigma $ in the vicinity of $C$. So it is interesting
to prove some theorems giving us information about the geometry of the
surface $\Sigma $ in the vicinity of $C$.

\textbf{Theorem}. \textit{The tangent plane to the surface $\Sigma$ at point
$C$ is the osculatory plane of $(\Gamma)$.}

\emph{Proof }

Let $\overrightarrow{OM}=\overrightarrow{\mathbf{F}(\tau )}$ the equation of
curve $(\Gamma )$ so that $\overrightarrow{{\mathbf{F}}(0)}$= $%
\overrightarrow{OC}$. The tangent plane at $M_{\alpha }$ or $M_{\beta }$ is
\begin{equation*}
\overrightarrow{M_{\alpha }M_{\beta }}=\overrightarrow{\mathbf{F}(\tau
_{\beta })}-\overrightarrow{\mathbf{F}(0)}-\overrightarrow{\mathbf{F}(\tau
_{\alpha })}+\overrightarrow{\mathbf{F}(0)}
\end{equation*}%
or, in the $(\Gamma )$ representation,
\begin{equation*}
\overrightarrow{M_{\alpha }M_{\beta }}=(\tau _{\beta }-\tau _{\alpha })%
\overrightarrow{\mathbf{F}^{\prime }(0)}+\frac{1}{2}\left( \tau _{\beta
}^{2}-\tau _{\alpha }^{2}\right) \overrightarrow{\mathbf{F}^{\prime \prime
}(0)}+...
\end{equation*}%
and
\begin{equation*}
\overrightarrow{\mathbf{F}^{\prime }(\tau _{\alpha })}=\overrightarrow{%
\mathbf{F}^{\prime }(0)}+\tau _{\alpha }\overrightarrow{\mathbf{F}^{\prime
\prime }(0)}+...
\end{equation*}%
It follows
\begin{equation*}
\overrightarrow{\mathbf{F}^{\prime }(\tau _{\alpha })}\wedge \overrightarrow{%
M_{\alpha }M_{\beta }}=\frac{1}{2}\left( \tau _{\alpha }-\tau _{\beta
}\right) ^{2}\overrightarrow{\mathbf{F}^{\prime }(0)}\wedge \overrightarrow{%
\mathbf{F}^{\prime \prime }(0)}+...
\end{equation*}%
When $\tau _{\alpha }$ and $\tau _{\beta }$ tends to $0$, the plane ($%
M_{\alpha },\overrightarrow{M_{\alpha }M_{\beta }},\overrightarrow{\mathbf{F}%
^{\prime }(\tau _{\alpha })}$) tends to the osculatory plane to $(\Gamma )$
at $C$ because the vector $\displaystyle\frac{2\,\overrightarrow{\mathbf{F}%
^{\prime }(\tau _{\alpha })}\wedge \overrightarrow{M_{\alpha }M_{\beta }}}{%
(\tau _{\alpha }-\tau _{\beta })^{2}}$ admits the limit $\overrightarrow{%
\mathbf{F}^{\prime }(0)}\wedge \overrightarrow{\mathbf{F}^{\prime \prime }(0)%
}$.

\textbf{Theorem}. \textit{The parametric representation of curve $\Gamma $
in Frenet frame at $C$ is}
\begin{equation*}
x=\tau \,,\,\,y=\tau ^{2}\,\,,\,\,z=\tau ^{4}\,\,;\quad \tau \in v(0)\,.
\end{equation*}%
\emph{Proof. }

Let {${C}$} an ordinary point of $(\Gamma )$. In a neighborhood of {${C}$},
we have
\begin{equation}
x=\tau \,,\,\,y=\tau ^{2}\,\,,\,\,z=\tau ^{3}\,\,;\quad \tau \in v(0)
\label{ordp}
\end{equation}%
such that $\tau =0$ corresponds to $C$. Let $M_{\alpha }$ and $M_{\beta }$
two points of $\Gamma $ with $\tau _{\alpha }\tau _{\beta }<0$. Eqs. (\ref%
{ordp}) lead to
\begin{equation}
\left( \overrightarrow{\mathbf{F}^{\prime }(\tau _{\alpha })}\wedge
\overrightarrow{M_{\alpha }M_{\beta }}\right) =\left(
\begin{array}{c}
-\tau _{\alpha }^{3}+2\tau _{\alpha }\tau _{\beta }^{2}-\tau _{\alpha
}^{2}\tau _{\beta } \\
2\tau _{\alpha }^{2}-\tau _{\beta }^{2}-\tau _{\alpha }\tau _{\beta } \\
\tau _{\beta }-\tau _{\alpha }
\end{array}%
\right)  \label{un}
\end{equation}%
and by changing $\tau _{\beta }$ with $\tau _{\alpha }$,
\begin{equation}
\left(\overrightarrow{\mathbf{F}^{\prime }(\tau _{\beta })}\wedge
\overrightarrow{M_{\beta }M_{\alpha }}\right) =\left(
\begin{array}{c}
-\tau _{\beta }^{3}+2\tau _{\beta }\tau _{\alpha }^{2}-\tau _{\alpha }\tau
_{\beta }^{2} \\
2\tau _{\beta }^{2}-\tau _{\alpha }^{2}-\tau _{\beta }\tau _{\alpha } \\
\tau _{\alpha }-\tau _{\beta }%
\end{array}%
\right) \,. \label{du}
\end{equation}%
The two vectors in Eqs.(\ref{un}, \ref{du}) are parallel if and only if
\begin{equation}
\left\{
\begin{tabular}{ll}
$\tau _{\alpha }^{2}+\tau _{\beta }^{2}-2\tau _{\alpha }\tau _{\beta }=0$ &
\\
$\tau _{\alpha }^{3}+\tau _{\beta }^{3}-\tau _{\alpha }\tau _{\beta
}^{2}-\tau _{\beta }\tau _{\alpha }^{2}=0.$ &
\end{tabular}%
\ \right.  \label{sys}
\end{equation}%
Equation (\ref{sys})$_{1}$ yields $\tau _{\alpha }=\tau _{\beta }$ and then (%
\ref{sys})$_{2}$ is verified. This means that any relation $\tau _{\beta
}=\phi (\tau _{\alpha })$ different from identity is not allowed and $\tau
_{\alpha }\tau _{\beta }<0$ is not possible. Then, the critical point {${C}$}
cannot be an ordinary point of $(\Gamma )$. Let us now consider the simplest
singular case for $\Gamma $ at $C$
\begin{equation*}
x=\tau \,\,,\,\,y=\tau ^{2}\,\,,\,\,z=\tau ^{4}\,.
\end{equation*}%
The parallelism conditions for vectors $\overrightarrow{\mathbf{F}^{\prime
}(\tau _{\alpha })}\wedge \overrightarrow{M_{\alpha }M_{\beta }}$ and $%
\overrightarrow{\mathbf{F}^{\prime }(\tau _{\beta })}\wedge \overrightarrow{%
M_{\beta }M_{\alpha }}$ are equivalent to
\begin{equation}
\left\{
\begin{tabular}{ll}
$-2\tau _{\alpha }^{4}-2\tau _{\alpha }^{3}\tau _{\beta }+2\tau _{\alpha
}^{2}\tau _{\beta }^{2}+2\tau _{\alpha }\tau _{\beta }^{3}=2\tau _{\beta
}^{4}+2\tau _{\beta }^{3}\tau _{\alpha }-2\tau _{\beta }^{2}\tau _{\alpha
}^{2}-2\tau _{\beta }\tau _{\alpha }^{3}$ &  \\
$3\tau _{\alpha }^{3}-\tau _{\alpha }^{2}\tau _{\beta }-\tau _{\alpha }\tau
_{\beta }^{2}-\tau _{\beta }^{3}=-3\tau _{\beta }^{3}+\tau _{\beta }^{2}\tau
_{\alpha }+\tau _{\beta }\tau _{\alpha }^{2}+\tau _{\alpha }^{3}$\,. &
\end{tabular}%
\ \right.  \label{syss}
\end{equation}%
Equation (\ref{syss})$_{2}$ implies $2(\tau _{\alpha }+\tau _{\beta
})(\tau _{\alpha }-\tau _{\beta })^{2}=0$ which has, for $\tau _{\alpha
}\neq \tau _{\beta }$, the unique possible solution $\tau _{\alpha }=-\tau
_{\beta }$ and then (\ref{syss})$_{1}$ is satisfied. \newline
The equations of the projection $(\Gamma ^{\prime })$ of the curve $(\Gamma
) $\ on the plane $y=\tau _{\alpha }^{2}$ are
\begin{equation*}
x=\tau ,\ \ z=\tau ^{4}
\end{equation*}%
and the straight line $M_{\alpha }M_{\beta }$ remains bitangent to $(\Gamma
^{\prime })$. \newline
Consequently, in the plane $y=\tau _{\alpha }^{2}$, the main part of the
equation of $(\Gamma ^{\prime })$ is written in the form
\begin{equation*}
z=H(\tau _{\alpha })\left( x-\tau _{\alpha }\right) ^{2}\left( x+\tau
_{\alpha }\right) ^{2}+\tau _{\alpha }^{4}\text{ \ \ with \ \ }H(0)\neq \ 0.
\end{equation*}%
In adimensional form, $H(0)=1$ and the main part of the equation of surface $%
\Sigma $ generated by this curve in a neighborhood of the critical point {${%
\ C}$}, when point $C$ is taken as origin of the axes and $Cxy$ is the
osculatory plane to $\Sigma $ at {${C}$, is
\begin{equation*}
z=(x^{2}-y)^{2}+y^{2}.
\end{equation*}%
The dimensions of $x$ and $y$ are different. In an adimensional system of axes, the equation writes%
\begin{equation*}
z=A\,\left\{ \left( B^{2}x^{2}-y\right) ^{2}+y^{2}\right\} .
\end{equation*}%
}

\end{document}